\documentclass[letterpaper]{article} % DO NOT CHANGE THIS
\usepackage{aaai23}  % DO NOT CHANGE THIS
\usepackage{times}  % DO NOT CHANGE THIS
\usepackage{helvet}  % DO NOT CHANGE THIS
\usepackage{courier}  % DO NOT CHANGE THIS
\usepackage[hyphens]{url}  % DO NOT CHANGE THIS
\usepackage{graphicx} % DO NOT CHANGE THIS
\usepackage{natbib}  % DO NOT CHANGE THIS AND DO NOT ADD ANY OPTIONS TO IT
\usepackage{caption} % DO NOT CHANGE THIS AND DO NOT ADD ANY OPTIONS TO IT
\usepackage{algorithm}
\usepackage{algorithmic}

\usepackage{newfloat}
\usepackage{listings}
\DeclareCaptionStyle{ruled}{labelfont=normalfont,labelsep=colon,strut=off} % DO NOT CHANGE THIS
\floatstyle{ruled}
\newfloat{listing}{tb}{lst}{}
\floatname{listing}{Listing}
\graphicspath{ {./images/} }
\usepackage{subcaption}
\usepackage{amssymb}

\title{Herd's Eye View: Improving Game AI Agent Learning with Collaborative Perception}
\author{
   Andrew Nash,\textsuperscript{\rm 1}
   Andrew Vardy, \textsuperscript{\rm 1 \rm 2}
   David Churchill \textsuperscript{\rm 1}
}
\affiliations{
    \textsuperscript{\rm 1} Department of Computer Science, Memorial University of Newfoundland, St. John's, Canada\\
    \textsuperscript{\rm 2} Department of Electrical and Computer Engineering, Memorial University of Newfoundland, St. John's, Canada\\
    anash@mun.ca, av@mun.ca, dave.churchill@gmail.com
}

\usepackage{bibentry}
\begin{document}

\maketitle

\begin{abstract}
We present a novel perception model named Herd's Eye View (HEV) that adopts a global perspective derived from multiple agents to boost the decision-making capabilities of reinforcement learning (RL) agents in multi-agent environments, specifically in the context of game AI. The HEV approach utilizes cooperative perception to empower RL agents with a global reasoning ability, enhancing their decision-making. We demonstrate the effectiveness of the HEV within simulated game environments and highlight its superior performance compared to traditional ego-centric perception models. This work contributes to cooperative perception and multi-agent reinforcement learning by offering a more realistic and efficient perspective for global coordination and decision-making within game environments. Moreover, our approach promotes broader AI applications beyond gaming by addressing constraints faced by AI in other fields such as robotics. The code is available on GitHub.\footnote{\url{https://github.com/andrewnash/Herds-Eye-View}}
\end{abstract}

\section{Introduction}
Game environments traditionally grant AI agents access to extensive global information from the game engine. While this configuration assists in efficient decision-making, it does not accurately represent the restrictions encountered by AI applications outside of gaming, where comprehensive access to a system's software or engine is not feasible. Consequently, game AI techniques that rely predominantly on game engine data may limit their potential contribution to broader AI applications, as their dependency on perfect information and global environmental data is often unrealistic in other contexts such as robotics and autonomous vehicles.

In response to these challenges, our work delves into the application of more constrained, realistic perception models for game AI. We take inspiration from publications like the ViZDoom platform \cite{Wydmuch_2019} and the Obstacle Tower Challenge \cite{juliani2019obstacle} that have embraced the shift towards game AI with real-world constraints. Both ViZDoom and Obstacle Tower have utilized visual data as the primary input for AI agents, enabling them to navigate complex 3D environments and reinforcing the importance of perception-based game AI models without game engine access.

Research in autonomous vehicles has made extensive strides in AI perception models, particularly using intermediary environmental representations like the Bird's Eye View (BEV). The BEV model provides an overhead perspective of the environment, often in the form of a semantic obstacle grid, from a single ``ego" vehicle's standpoint. This concept has become a key component in many self-driving systems \cite{Ma2022VisionCentricBP}.

Drawing on these past works, we propose a similar intermediary representation for game AI: the Herd's Eye View (HEV) model. Differing from the BEV's ego-centric perspective, the HEV model offers a shared world-centric perception derived from multiple agents. This shared perception model aligns closer to real-world AI applications, where multiple systems often work together to understand and navigate their environment.

The HEV model presents dual advantages. First, it mirrors the constraints faced by AI outside of gaming, contributing to the development of more believable AI behavior in games. Second, it alleviates the computational demands associated with the BEV model, where each agent maintains its own unique view of the environment, instead, only a single shared global view is utilized.

Emulating the successful methodologies of the ViZDoom project and the Obstacle Tower paper, we also incorporate Reinforcement Learning (RL) into our approach. RL enables us to test the effectiveness of HEV in both low-level control tasks and high-level planning challenges concurrently in complex environments. Importantly, similar to the Obstacle Tower approach, our agents are assessed not solely on their ability to navigate familiar environments, but also on their ability to handle unique variations of these environments. This highlights the importance of generalization in adapting to novel scenarios within the same environment.

To assess the effectiveness of the HEV model, we conduct two sets of experiments in three simulated Multi-Agent Reinforcement Learning (MARL) game environments. The first compares the accuracy of HEV world-centric predictions with BEV ego-centric predictions. The second experiment evaluates the efficiency of policies learned by RL agents trained on HEV perspective views compared to those trained on BEV perspective views. 

Our work makes the following contributions:
\begin{enumerate}
\item We propose a baseline model for performing semantic segmentation in a fixed ``HEV'' world-centric view.
\item We demonstrate the effectiveness of the HEV fixed world viewpoint in improving collaborative perception and MARL in games.
\end{enumerate}

Our exploration of more realistic perception models provides significant insights for game AI development, stressing the wider applicability of these techniques beyond the gaming industry.

\section{Related Works}

\subsection{Birds Eye View Semantic Segmentation}
In autonomous vehicle research, the bird's-eye view semantic segmentation task involves predicting pixel-level semantic labels for a top-down ego-centric view of an environment. Segmentation classes are typically dedicated to vehicles, driveable areas, and obstacles. In prior BEV research, a significant point of distinction lies in the method used for transforming 2D perspective-view features into 3D space or directly onto the BEV plane. Many previous works have leveraged explicit geometric reasoning in their perspective transformation \cite{ReiherLampe2020Cam2BEV, philion2020lift, fiery2021}. An approach that has recently gained popularity is the Cross-View Transformer (CVT) \cite{zhou2022cross} model, which implicitly models scene geometry. The CVT model leverages a camera-aware cross-view attention mechanism to implicitly learn a mapping from individual camera views to a canonical map-view representation for map-view semantic segmentation. The model consists of a convolutional image encoder for each view and cross-view transformer layers for inferring the segmentation in a simple, easily parallelizable, and real-time manner. BEVFormer \cite{li2022bevformer} uses a similar cross-attention model to extract spatiotemporal BEV information. BEVSegFormer \cite{peng2022bevsegformer} uses a deformable transformer-based encoder. There are many publications in this research area using similar architectures of transformers to shift perspective view(s) to BEV, \citeauthor{Ma2022VisionCentricBP} provides a recent review of these architectures.

% \begin{figure}[!t]
%   \centering
%   \includegraphics[width=\linewidth]{POVDiagram}
%   \caption{A visualization of the different map-view coordinate frames tested, ego-centric views move in accordance to translation and rotation of the agent, whereas the world-centric view does not.}
%   \label{fig:bev_pov}
% \end{figure}

The HEV semantic segmentation task poses a unique challenge compared to the BEV task since the agent translations are unknown; this requires the model to geometrically reason about multiple camera views to localize. For our baseline approach, we leverage the CVT model proposed by \cite{zhou2022cross}. The CVT model is well suited for the HEV task because of its global attention mechanism. Many BEV publications such as BEVFormer \cite{li2022bevformer} and BEVSegFormer \cite{peng2022bevsegformer} aim to optimize this global attention mechanism since in ego-centric tasks, a camera view only overlaps with a consistent subsection of the map-view. Conversely, in our HEV world-centric use case, global attention is an advantage because a camera view can overlap with any part of the map-view. Additionally, we expect that the model's performance can be further improved by incorporating additional information from other sensors, such as lidar and radar, as demonstrated by recent works \cite{harley2022simple}.

\subsection{Collaborative Perception Datasets}

\begin{figure*}[!t]
  \centering
  \includegraphics[width=\linewidth]{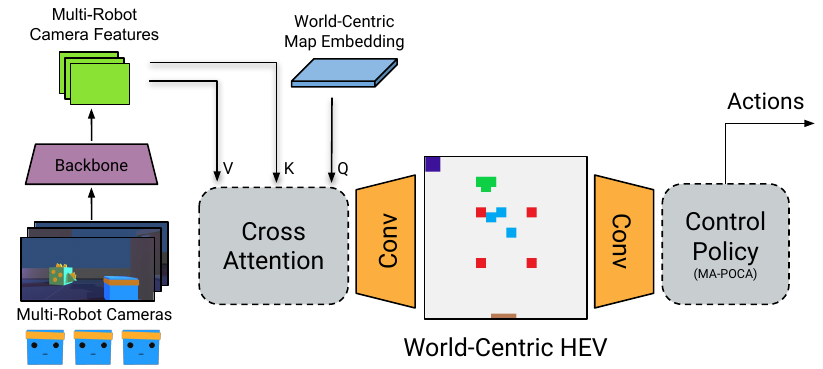}
  \caption{A visualization of the proposed HEV approach in the Dungeon Escape environment: Agent camera views are extracted via a backbone model, then combined in a cross-attention module, then decoded into a world-centric semantic segmentation. The resulting semantic segmentation can be used as an observation for a swarm of robots.}
  \label{fig:hev_method}
\end{figure*}

Autonomous vehicle datasets have been widely used in collaborative perception research, comprising various sensory inputs, including cameras, lidar, and GPS \cite{Han2023CollaborativePI}, from multiple vehicles in a vehicle-to-vehicle environment \cite{xu2022opencood, yu2022dairv2x}. Some datasets, such as those proposed in \cite{li2022v2x, dolphins2022}, include infrastructure sensors, resulting in a vehicle-to-infrastructure data model. Others, such as the dataset presented in \cite{xu2022v2xvit}, employ a vehicle-to-everything model. The CoPerception-UAVs dataset \cite{Where2comm:22} employs five images from five drones flying in formation. It is worth noting that these datasets are all sourced from CARLA \cite{Dosovitskiy17} in Unreal Engine, a widely used open-source platform for simulating autonomous vehicles.

The HEV datasets sourced from our simulated environments are uniquely challenging in the field of collaborative perception, as the agents are equipped with only one or two cameras. Unlike previously proposed collaborative perception problems, the HEV task does not provide the agents with the transformation component of their pose. The unknown position of each camera view within the global coordinate frame adds a significant challenge to the semantic segmentation prediction task and other downstream tasks.

\subsection{Collaborative Perception Methods}

Collaborative perception has been explored in recent years, improving the capability of single-agent perception models \cite{li2022multi, Where2comm:22, LACP, Su2022uncertainty, zhou2022multi}. In conventional collaborative perception, intermediate representations produced by sensors or neural networks from multiple viewpoints are propagated among a team of robots, such as a group of vehicles \cite{Where2comm:22, xu2022cobevt} or a swarm of drones \cite{zhou2022multi, Where2comm:22}. The existing works commonly learn a collaboration module, produced by a Graph Neural Network \cite{zhou2022multi, Li_2021_NeurIPS}, Convolutional Neural Network \cite{li2022multi, Qiao_2023_WACV}, or a Transformer \cite{xu2022cobevt, Where2comm:22} to combine multiple robot intermediate representations. 

Prior research has focused on robots equipped with multiple sensors, requiring sensor data fusion on a per-agent basis before information exchange among agents \cite{Han2023CollaborativePI}. However, in this work, we focus on robots with only one or two cameras and no additional sensors, making our approach more amenable to smaller, simpler robot swarms. Since we focus on simpler robots, we do not utilize a collaboration module, and instead fuse all camera views together in a single cross-attention module.

\section{Methodology}

% \begin{figure*}[!t]
%     \begin{subfigure}[b]{0.33\textwidth}
%             \includegraphics[width=\linewidth]{PushBlock}
%             \caption{}
%             \label{fig:pushblock}
%     \end{subfigure}%
%     \begin{subfigure}[b]{0.33\textwidth}
%             \includegraphics[width=\linewidth]{DungeonEscape}
%             \caption{}
%             \label{fig:dungeonescape}
%     \end{subfigure}%
%     \begin{subfigure}[b]{0.33\textwidth}
%             \includegraphics[width=\linewidth]{PlanarConstruction}
%             \caption{}
%             \label{fig:planarconstruction}
%     \end{subfigure}
%     \caption{Images of the three environments used to test the HEV collaborative perception and reinforcement learning algorithms. From Left-to-right: (a) Collaborative Push Block, (b) Dungeon Escape, (c) Planar Construction}\label{fig:envs}
% \end{figure*}

\subsection{Herd's Eye View}
 In the Herd's Eye View (HEV) semantic segmentation task, we are given a set of $n$ monocular camera views, $(I_k, K_k, R_k)^n_{k=1}$ consisting of an input image $I_k \in \mathbb{R} ^{H \times W \times 3}$, camera intrinsics $K_k \in \mathbb{R} ^{3 \times 3}$, and extrinsic rotation $R_k \in \mathbb{R} ^{3 \times 3}$ with respect to the agent base. The goal of the HEV task is to predict a binary semantic segmentation mask  $y \in \{0, 1\} ^{h \times w \times C}$ in the global coordinate frame, where $C$ is the desired number of segmentation classes. The HEV task adds additional ambiguity to the well-studied BEV task as each camera view is at an unknown translation and orientation with respect to the global coordinate frame.

We define a BEV as a single-agent perception transformed into an ego-centric view, whereas the HEV is a collaborative perception transformed into a fixed world-centric view. A comparison of the ego-centric views tested and the fixed word-centric view can be seen in Figure \ref{fig:pov_push_block}.

Our approach, seen in Figure \ref{fig:hev_method}, follows three steps:
\begin{enumerate}
    \item Collect multiple views of the environment from robot cameras. 
    \item Use a collaborative perception model to obtain the HEV, the world-centric semantic segmentation of the environment.
    \item Input the HEV to a Reinforcement Learning (RL) control policy to obtain agent control commands. 
\end{enumerate}

Our goal is to establish a baseline HEV perception model to extract information from the multiple camera views and project them onto a fixed world-centric view. We propose a baseline perception model using the Cross-View Transformer (CVT) \cite{zhou2022cross} and use semantic segmentation as our downstream task. The Cross-View Transformer is a recent approach that uses a cross-view attention module, first proposed by \cite{zhou2022cross}, enabling the agents to reason about their environment in an ego-centric coordinate frame. We extend the CVT model to further improve its accuracy and speed for the HEV use case. We name our baseline model the Herd's Eye View Cross-View Transformer (HEV-CVT). We use a world-centric map embedding and tune positional embeddings, output sizes, and the number of transformer layers to fit our proposed HEV environments.

\subsection{Data Collection}

\begin{figure*}[!t]
  \centering
  \includegraphics[width=0.8\linewidth]{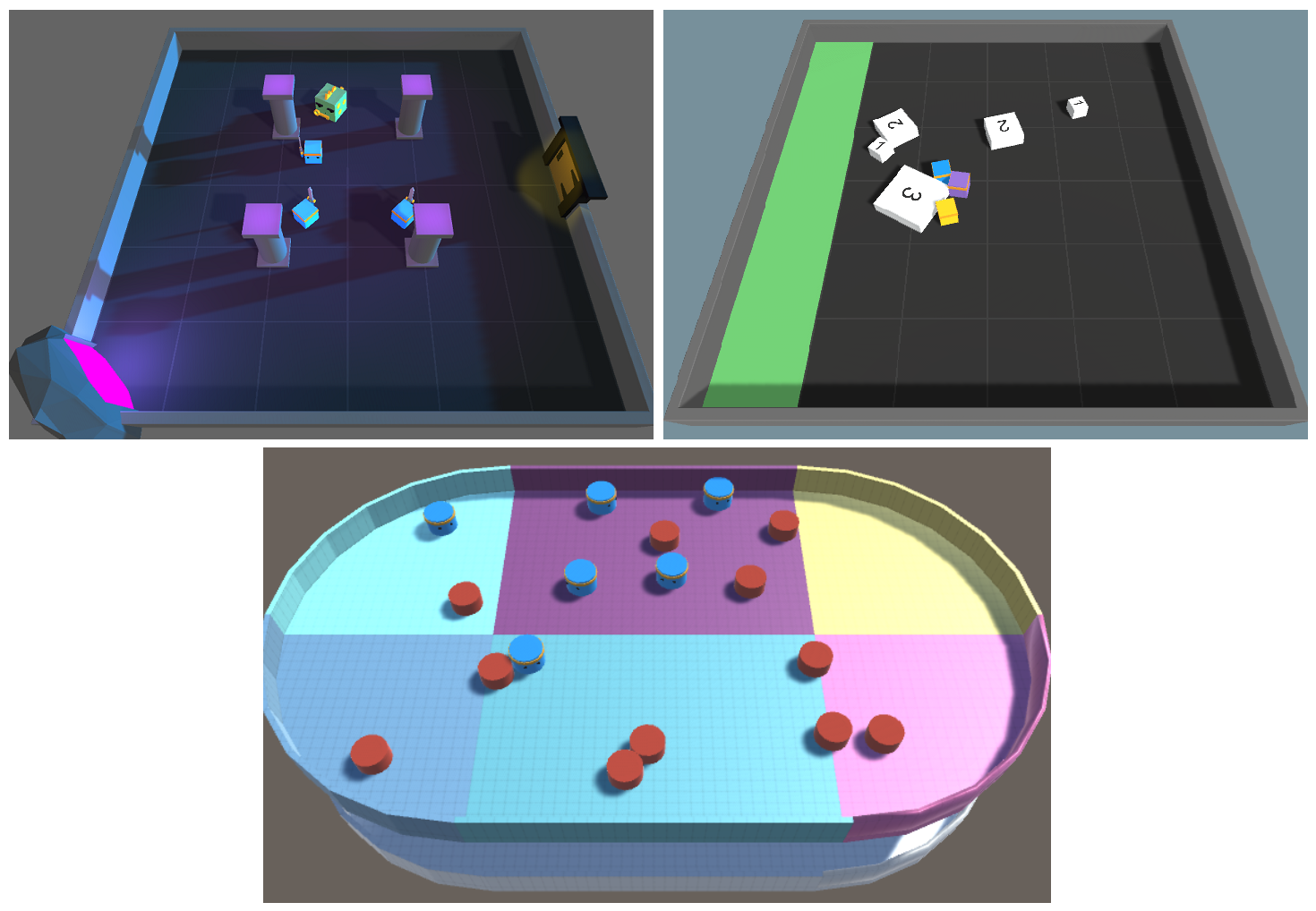}
  \caption{Images of the three environments used to test the HEV collaborative perception and reinforcement learning algorithms. The top left is the Dungeon Escape environment. The top right is the Collaborative Push Block environment. The bottom is the Planar Construction environment.}
  \label{fig:envs}
\end{figure*}

\begin{figure*}[!t]
  \centering
  \includegraphics[width=0.83\linewidth]{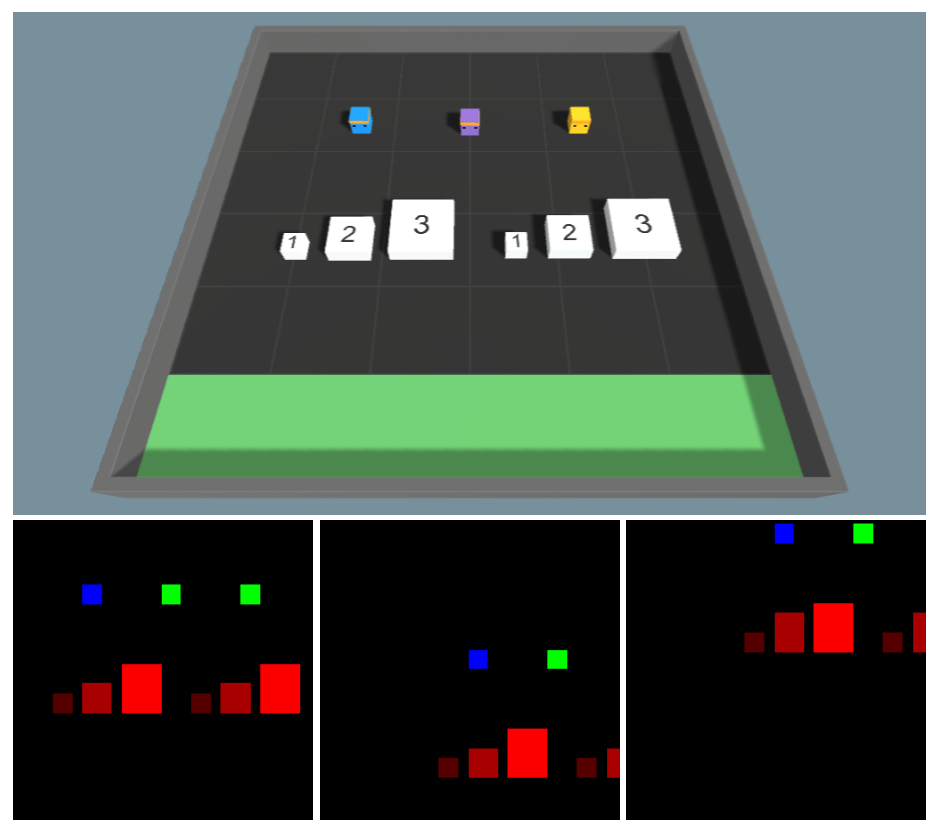}
  \caption{Example scene and corresponding agent observations from the Collaborative Push Block environment. The top image shows a debug camera (not available for agent observation). The bottom left shows the HEV world-centric observation of the blue agent. The bottom middle shows the BEV-centric observation of the blue agent. The bottom right shows the BEV-forward observation of the blue agent. Blue is the controller agent, green is ally agents, and red shades are differently-sized push blocks.}
  \label{fig:pov_push_block}
\end{figure*}

We use identical Unity simulation environments to source the datasets for training the HEV semantic segmentation task and MARL task. To collect the HEV ground truth for both tasks, we use our own custom fork of MBaske's Unity Grid-Sensor Library \cite{gridsensor} which allows the collection of HEV world-centric grid-sensors. The only difference between ego-centric based agents and world-centric based agents is the location of their grid-sensor and the perspective at which they take their actions (e.g., forward for the word-centric agent is always North, but forward for the ego-centric agent is with respect to their current orientation). All agents are trained on ground-truth sensors, calculated using the bounding boxes that are individually tuned to each object. The resolution of the grid-sensor is adjusted to accommodate the complexity and size of the environment as seen in Table \ref{table:RL_env_params}. Example observations of world-centric and ego-centric based agents can be seen in Figure \ref{fig:pov_push_block}

Our simulations are conducted in three different Unity ML-Agents environments:

\textbf{Collaborative Push Block:} Three agents are required to push white blocks to a green goal area on a randomly selected side of a square area. There are blocks sized one, two and three, each requiring the respective amount of agents to push into the goal area \cite{cohen2022}.

\textbf{Dungeon Escape:} As a Green Dragon slowly walks towards an exit portal, one of the three agents must collide with it in order to sacrifice itself and spawn a key. The key must then be picked up by one of the two remaining agents and brought to the exit door to escape the dungeon \cite{cohen2022}. Once any agent escapes, all agents win.

\textbf{Planar Construction:} Six agents collaborate to push red pucks into desired positions. Desired positions are randomly assigned to valid coordinates within the arena, and are observed via a Grid-Sensor, similar to the Push Block environment \cite{strickland19rl}. In each round a new random amount of pucks from 2 to 16 are spawned.

We utilize the open-source Collaborative Push Block and Dungeon Escape environments from ML-Agents \cite{cohen2022}, which are already native to Unity and only change the sensor input of agents. We recreate the Planar construction task \cite{VardyOC, strickland19rl, vardy2020distancing, vardy2022lasso, vardy23labyrinth} based on \citeauthor{strickland19rl}'s work in the CWaggle simulator but adapt the environment to Unity ML-Agents. All three environments can be seen in Figure \ref{fig:envs}. For MARL training, we use the HEV ground truth as model input and identical reward functions to the original implementations. Specifically, the agents are trained using the Multi-Agent POsthumous Credit Assignment (MA-POCA) algorithm \cite{cohen2022} in Unity ML-Agents. By using identical reward functions, we aim to create a fair comparison between the performance of agents using HEV and those using traditional sensor frames in cooperative scenarios.

The MARL task enables us to train the CVT models, which can perform semantic segmentation in an ego-centric or world-centric view. To collect the data necessary for training the CVT models, we run the trained MA-POCA models and collect the camera view, camera intrinsics, and rotation extrinsic from each agent at each step of the simulation, along with the ground truth HEV and BEV. By collecting data from various environments and introducing variations, we aim to create diverse and robust datasets for training the CVT models.

\subsection{Implementation Details}

\begin{figure*}[!t]
  \centering
  \includegraphics[width=\linewidth]{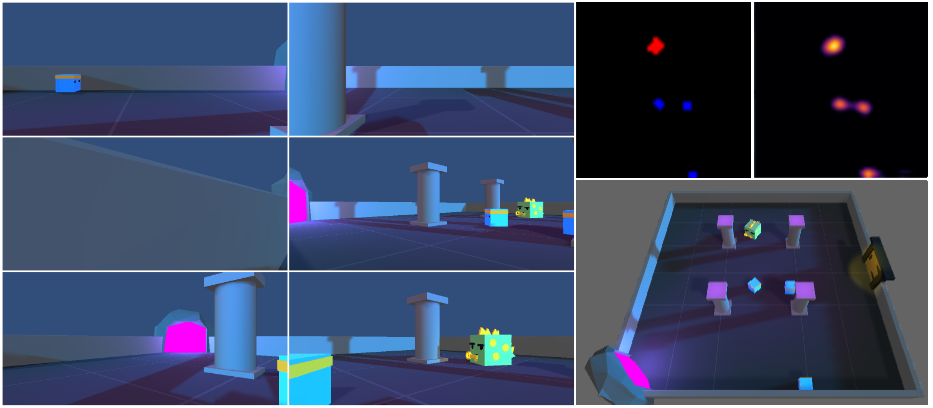}
  \caption{Sample HEV-CVT prediction from the Dungeon Escape environment validation dataset. The two left columns show each of the three agents' unique camera views, each row contains the images from the left and right cameras of the same agent. The top right shows the HEV-CVT prediction confidence heat map of the agents and dragon, the ground truth is directly to the left (agents are blue, the dragon is red). The bottom right shows a world-view camera not available to agents to help readers understand the scene.}
  \label{fig:hev_sample_pred}
\end{figure*}

\begin{table*}[!t]
    \renewcommand{\arraystretch}{1.5}
    \caption{HEV simulated environment parameters.}
    \label{table:RL_env_params}
    \centering
    \begin{tabular}{|l|l|l|l|}
        \hline 
         & Agents & Agent Cameras & Grid Size \\
        \hline
        Collaborative Push Block & 3    & 1-left, 1-right   & 32x32 \\
        \hline
        Dungeon Escape           & 2-3  & 1-left, 1-right   & 32x32 \\
        \hline
        Planar Construction      & 6    & 1-forward         & 32x64 \\
        \hline
    \end{tabular}
\end{table*}

The Cross-View Transformer is adapted from \citeauthor{zhou2022cross} for the Herds Eye View Collaborative Perception task. The first stage of the network passes each input image from agents into a feature extractor, we use an EfficientNet-B4 \cite{DBLP:journals/corr/abs-1905-11946}, which outputs two multi-resolution patch embeddings of size (28, 60) and (14, 30). Each patch is passed into a Cross-View Transformer convolution stack as in the original implementation. We found fewer convolution stacks significantly degrade the HEV-CVTs ability to localize, and more are not necessary. The patch embedding act as image features and are used in the keys and as the values for the Cross-View Transformer.

We encode the rotation $R_k \in \mathbb{R} ^{3 \times 3}$ of the agent's camera into a $D$-Dimensional positional embedding using a multilayer perceptron. We use $D = 64$ for all of our experiments. The positional embedding is combined with the image feature to compute the keys for the cross-view transformer. The world-centric map embedding operates similarly to the originally proposed map-view embedding. The key difference with our approach is we do not subtract camera location embeddings from the map embedding, instead, we directly use the learned map embedding as queries. The camera locations with respect to the world are unknown for the HEV task, and we found subtracting rotation embeddings did not improve performance. The transformer architecture refines its world-centric estimate through two rounds of computation, each resulting in new latent embeddings used as queries.

The cross-view transformer computes softmax-cross-attention \cite{attentionisallyouneed} between the image feature keys, values and world-centric queries. This setup allows world coordinates from the world-centric map embedding to attend to one or more image locations, allowing the model to reason about the environment from multiple image features. The multi-head attention mechanism uses 4-heads like the original implementation but with half the embedding size of $d_{head}=32$.

The cross-view transformer output is 8x8 for square environments and 8x8 and 8x16 for rectangular environments, this then passes through a decoder consisting of three up-convolutional layers to a final size of 64x64 and 64x128. This is purposely larger than is required for RL observation size, as smaller sizes can create ambiguity for some object occupancy resulting in decreased performance. These larger HEV-CVT sizes can easily be down-sampled to match the required RL observation sizes of 32x32 and 32x64. We threshold the output prediction confidences, keeping predictions with a confidence greater than 0.4. The prediction confidences prior to thresholding can be seen in Figure \ref{fig:hev_sample_pred} as a heat map (lighter is higher confidence).

Our training process is similar to the original implementation by \citeauthor{zhou2022cross}, we also use focal loss \cite{Lin2017FocalLF} and the AdamW \cite{DBLP:journals/corr/abs-1711-05101} optimizer with a one-cycle learning rate scheduler \cite{smith2018superconvergence}. All models are trained with a batch size of 4 for 25 epochs. Training lasts approximately 8 hours on a single RTX 3090 GPU before converging.

\section{Experiments and Results}

\begin{table*}[t]
    \renewcommand{\arraystretch}{1.5}
    % if using array.sty, it might be a good idea to tweak the value of
    % \extrarowheight as needed to properly center the text within the cells
    \caption{HEV-CVT validation IoU results per coordinate frame in each environment (higher is better).}
    \label{table:perception_results}
    \centering
    % Some packages, such as MDW tools, offer better commands for making tables
    % than the plain LaTeX2e tabular which is used here.
    \begin{tabular}{|l|l|l|l|}
        \hline
         & World-Centric & Ego-Centric & Ego-Forward \\
        \hline
        Collaborative Push Block & \textbf{96.94\%} & 63.87\% & 64.22\% \\
        \hline
        Dungeon Escape           & \textbf{43.53\%} & 13.47\%  & 26.07\% \\
        \hline
        Planar Construction      & \textbf{48.37\%} & 35.45\% & 10.16\% \\
        \hline
    \end{tabular}
\end{table*}

\begin{table*}[!t]
    \renewcommand{\arraystretch}{1.5}
    \caption{MA-POCA mean episode length $\pm$ standard deviation per coordinate frame in each environment (lower is better).}
    \label{table:RL_results}
    \centering
    \begin{tabular}{|l|l|l|l|}
        \hline
          & World-Centric & Ego-Centric & Ego-Forward \\
        \hline
        Collaborative Push Block & \textbf{100.1} $\pm$ 40.6 & 137.9 $\pm$ 53.5 & 124.9 $\pm$ 47.4 \\
        \hline
        Dungeon Escape           & \textbf{15.1}  $\pm$ 0.81  & 17.3 $\pm$ 0.87 & 18.4 $\pm$ 1.44 \\
        \hline
        Planar Construction      & \textbf{176.9} $\pm$ 40.8 & 233.8 $\pm$ 73.7 & 239.8 $\pm$ 75.8 \\
        \hline
    \end{tabular}
\end{table*}

\subsection{Collaborative Perception}

The HEV-CVT model must accurately localize the position of each agent based on the overlap of camera frames, which are located at unknown positions. An example of this in the Dungeon Escape environment can be seen in Figure \ref{fig:hev_sample_pred}. The cameras are recorded at resolution $480 \times 224$, and we use the camera intrinsics of a Raspberry Pi Camera Module 3. Consistent with prior works \cite{Ma2022VisionCentricBP}, we show the result Intersection over Union (IoU) metric for the HEV-CVT model trained on each environment in Table \ref{table:perception_results}. We compare the performance of the baseline CVT model on world-centric, ego-centric, and ego-forward coordinate frames. 

In the Collaborative Push Block environment, three agents are equipped with two forward-facing cameras and are tasked with predicting the occupancy of all push blocks, agents and the goal area. In the Dungeon Escape environment, three agents are equipped with two forward-facing cameras and are tasked with predicting the occupancy of the dragon, agents and key. In the Planar Construction environment, six agents are equipped with a single forward-facing camera and are tasked with predicting the occupancy of all pucks and agents.

Our results shown in Table \ref{table:perception_results} demonstrate the world-centric coordinate frame consistently outperforms the ego-centric coordinate frames in all environments. The Collaborative Push Block and Dungeon Escape environments show the largest performance improvements, with up to 32.72\% and 17.46\% improvement in IoU, respectively. These results suggest that the world-centric HEV approach is effective in addressing the challenges of collaborative perception in multi-agent environments. This result is especially apparent in the Collaborative Push Block environment, where the HEV-CVT model easily localizes itself based on the large goal location seen in most camera views for a near-perfect 96.94\% IoU score. The landmarks in the Dungeon Escape environment, the exit door and portal are in randomized locations which makes localization harder than the Push Block environment, reflected by the steep drop in IoU scores.

The standard ML-Agents environments were not as challenging for the CVT models as there were not many permutations of the environment layout. By contrast, our custom Planar Construction environment presents a more complex challenge as we randomly change the coloring of six wall and floor components at every time step of the environment during data collection. Additionally, the locations of pucks to be pushed are randomized, and the environment area is twice the size of the ML-Agents environments. Despite the additional challenge the HEV-CVT model still performs well in the Planar Construction environment scoring 48.37\% on the HEV task. This result shows the CVT models can localize based on the overlap in views between cameras as much of the validation set contains wall colors and puck layouts never before seen.

\subsection{Multi-Agent Reinforcement Learning}

In order to compare the performance of the fixed world-centric coordinate frames with other commonly used coordinate frames, we conduct experiments in all three proposed environments. To ensure a fair comparison between the performance of agents using different coordinate frames, we use identical reward functions to each environment's original implementation and identical grid sizes.

Table \ref{table:RL_results} compares the performance of agents using different coordinate frames in all three proposed environments. We find consistently lower episode lengths with world-centric based agents compared to ego-centric. We opt to use episode length as our performance metric, as it directly reflects the speed of task completion. While alternative metrics such as cumulative or mean reward are also commonly used, these primarily reflect minor negative rewards assigned per time step, providing less insight into an agent's efficiency in our context.

Our experiments highlight a common challenge faced by BEV-based agents in all three environments, often an object necessary to take the optimal action was missing from the agent's view, leading to sub-optimal decision-making and increased episode lengths. This was especially apparent in the Push Block environment where often one of the three agents would not observe the size three block (requiring all three agents to push it), causing two agents to be waiting for the third agent to join them at the block, wasting time. Conversely, we found HEV-based agents in the Push Block environment stuck close together and consistently pushed the highest value blocks together first.

The HEV-based agents were able to leverage the multiple viewpoints available to them, enabling them to better perceive their environment and take more optimal actions. This issue was particularly evident in the Push Block environment, where the improved perception of world-centric agents resulted in significantly lower episode lengths than ego-based agents.

Overall, these findings suggest that the HEV framework offers a superior perception model in MARL environments, providing agents with a more comprehensive understanding of their surroundings, leading to improved decision-making and better overall performance.

\section{Conclusion}
We have proposed a new perception model called Herd's Eye View that provides a global view of the environment, enabling better global coordination and cooperation in MARL scenarios. We conduct two sets of experiments in three simulated multi-agent environments. Our first experiment focuses on the perception aspect of HEV and shows the same Cross-View Transformer model performs better on the world-centric HEV task than its BEV ego-centric counterpart. Our second experiment focuses on the effectiveness of the HEV perspective view compared to BEV perspective views for MARL agents. We find that RL agents trained on world-centric perspective views learn more efficient policies than those trained on ego-centric perspective views. Our work opens up new possibilities for advanced perception models in MARL game environments, which can greatly enhance the performance of multi-agent systems by enabling better collaboration and coordination.

\bigskip

\bibliography{aaai23}

\end{document}